\documentclass[english,english,noshowpacs, amssymb,twocolumn,aps,nofootinbib]{revtex4}
\usepackage[T1]{fontenc}
\usepackage[latin9]{inputenc}
\usepackage{color}
\usepackage{mathtools}
\usepackage{textcomp}
\usepackage{amsmath}
\usepackage{esint}
\usepackage{url}
\usepackage{natbib}
\usepackage{graphicx}
\usepackage[export]{adjustbox}

\makeatletter

\newcommand{\lyxmathsym}[1]{\ifmmode\begingroup\def\b@ld{bold}
  \text{\ifx\math@version\b@ld\bfseries\fi#1}\endgroup\else#1\fi}

\@ifundefined{textcolor}{}
{%
 \definecolor{BLACK}{gray}{0}
 \definecolor{WHITE}{gray}{1}
 \definecolor{RED}{rgb}{1,0,0}
 \definecolor{GREEN}{rgb}{0,1,0}
 \definecolor{BLUE}{rgb}{0,0,1}
 \definecolor{CYAN}{cmyk}{1,0,0,0}
 \definecolor{MAGENTA}{cmyk}{0,1,0,0}
 \definecolor{YELLOW}{cmyk}{0,0,1,0}
}
\makeatother

\usepackage{babel}
\begin{document}

\title{Quantum-Enhanced Plasmonic Sensing}

\author{Mohammadjavad~Dowran$^{1,*}$, Ashok~Kumar$^{1,}${{\footnote {These authors contributed equally to this work.}}}, Benjamin~J.~Lawrie$^{2}$, Raphael~C.~Pooser$^{2}$, Alberto~M.~Marino$^{1,}${\footnote {marino@ou.edu}}}

\affiliation{$^{1}$Homer L. Dodge Department of Physics and Astronomy, The University of Oklahoma, Norman, Oklahoma 73019, USA
\\
$^{2}$Quantum Information Science Group, Oak Ridge National Laboratory, Oak Ridge, Tennessee 37831, USA}





\begin{abstract}
Quantum resources can enhance the sensitivity of a device beyond the classical shot noise limit and, as a result, revolutionize the field of metrology through the development of quantum-enhanced sensors.
In particular, plasmonic sensors, which are widely used in biological and chemical sensing applications, offer a unique opportunity to bring such an enhancement to real-life devices.
Here, we use bright entangled twin beams to enhance the sensitivity of a plasmonic sensor used to measure local changes in refractive index.
We demonstrate a 56\% quantum enhancement in the sensitivity of state-of-the-art plasmonic sensor with measured sensitivities on the order of~$10^{-10}$~RIU$/\sqrt{\textrm{Hz}}$, nearly 5 orders of magnitude better than previous proof-of-principle implementations of quantum-enhanced plasmonic sensors.
These results promise significant enhancements in ultratrace label free plasmonic sensing and will find their way into areas ranging from biomedical applications to chemical detection.
\end{abstract}


\maketitle

\section{Introduction}

The field of quantum metrology takes advantage of non-classical resources to enhance the precision and resolving power of classical measurement techniques beyond the shot noise limit (SNL)~\cite{Giovannetti,taylor2016quantum}. In particular, for optical based sensing, this limit defines the minimum noise floor achievable with classical resources and can only be surpassed through the use of quantum states of light such as squeezed or entangled states~\cite{Walls,Slusher}. These states allow for a lower noise floor, which makes it possible to detect signals below the SNL, thus increasing the resolving power of the sensor. This has led to a number of proposals that seek to use these states to enhance classical techniques such as interferometry, spectroscopy, spatial based measurements, and imaging~\cite{Caves,Xiao,Grangier,Collaboration,Taylor,Treps,Brida}.

While the use of quantum states of light to enhance measurements was originally proposed about three decades ago, experimental realizations of quantum-enhanced sensors are only beginning to emerge. Furthermore, the interface between plasmonic sensors and quantum states of light offers a unique opportunity to bring quantum-based sensitivity enhancements to devices that are already used in real-life applications. Due to their robust diagnostic capabilities, plasmonic sensors have found their way into a number of applications such as bio-sensing, atmospheric monitoring, ultrasound diagnostics, and chemical detection~\cite{NatMat,Homola2008,Homola,Gordon,Schuller,Yakovlev,Stewart}. In addition, these sensors rely on optical readout techniques and already operate at the SNL~\cite{Wang}. Thus, further enhancements for a particular geometry are only possible through the use of quantum states of light.

Here, we demonstrate the ability of quantum states of light to enhance the sensitivity of state-of-the-art plasmonic sensors. While there have been proof-of-principle experimental and theoretical studies of quantum-enhanced plasmonic sensors~\cite{Kalashnikov,Pooser,LawriePooser,Lee,JSLee}, the results presented here represent the first implementation of such a sensor with sensitivity of the same order of magnitude as the classical state-of-the-art. In particular, we detect changes in refractive index of air induced by ultrasonic waves.
This practical implementation of quantum-enhanced sensing can be easily extended to other measurement configurations and paves the way for real-life applications.
More significantly, the level of sensitivity achieved here represents an improvement  of nearly five orders of magnitude with respect to previous proof-of-principle implementations of quantum-enhanced plasmonic sensors~\cite{Kalashnikov,Pooser,LawriePooser} and of over two orders of magnitude with respect to the classical state-of-the-art ultrasound sensing with plasmonic sensors~\cite{Yakovlev}.

We consider surface plasmon resonance (SPR) sensors that consist of an array of sub-wavelength nanostructured holes in a thin silver film. The operation of these sensors is based on the optical excitation of electron oscillations at the interface between a metal and a dielectric, or surface plasmons. This coherent conversion between photons and plasmons gives rise to a transmission through the sub-wavelength holes orders of magnitude greater than transmissions expected from diffraction theory, an effect known as extraordinary optical transmission (EOT)~\cite{Ebbesen,Martin-Moreno}. This process preserves the quantum properties of the light~\cite{Altewischer,Fasel,Lawrie,Holtfrerich} and makes the use of quantum states of light a viable option to enhance the sensitivity of plasmonic sensors~\cite{Lee,JSLee}.

The characteristics of the plasmon resonance are determined by the shape, periodicity, and dimensions of the nano-holes in the array, the thickness of the metal film, and the refractive index of the metal and surrounding dielectric. As a result, SPR sensors exhibit a frequency shift of their resonant response and a corresponding change in optical transmission at a given wavelength as a result of small changes in local refractive index~\cite{Stewart}, which can be caused by the selective binding of the analyte of interest or a change in pressure.

In general, for EOT sensors, intensity ($I$) measurements are used to estimate the refractive index ($n$). In this case, the sensitivity, or minimum resolvable change in the index of refraction ($\Delta n_{\rm min}$), is given by:
\begin{equation}
      \Delta n_{\rm min}=\frac{1}{|\partial T/\partial n|}\frac{\sqrt{(\Delta I)^2}}{|\partial I/\partial T|}.
\end{equation}
Here, the term $|\partial T/\partial n|$ characterizes the change in transmission ($T$) with a change in refractive index and is determined by the properties of the plasmonic structure, while the term $\sqrt{(\Delta I)^2}/|\partial I/\partial T|$ characterizes the properties of the optical field used to probe the sensor~\cite{Piliarik}. This points to different mechanisms to obtain a sensitivity enhancement. On the one hand, the sensitivity can be increased by modifying the response of the plasmonic structure to changes in refractive index, for which several techniques have been proposed~\cite{Guo}. On the other hand, an enhancement can be obtained through an increase in intensity or a reduction of the noise properties of the light used to probe the sensor.  However,  an increase in probing power is not always an option due to the optical damage threshold of the substance under study, such as biological samples~\cite{Taylor,NatMat}, or the plasmonic structures themselves~\cite{Baffou}.

\section{Experiment}

In our experiment, we probe the plasmonic sensors with one of the entangled twin beams of light generated via a four-wave mixing (FWM) process based on a double-$\Lambda$ configuration in hot $^{85}$Rb atoms~\cite{FWM-Lett}, as shown in Fig.~1. To implement the FWM, a strong pump beam (power of 550 mW and $1/e^{2}$ waist diameter of 1.0~mm) and a 3.04~GHz frequency down-shifted probe beam ($1/e^{2}$ waist diameter of 0.7~mm) are derived from a Ti-Sapphire laser operating at a wavelength of 795~nm. These two orthogonally polarized beams intersect at an angle of 0.5~degrees at the center of a 12~mm long $^{85}$Rb vapor cell heated to 109$^\circ$C, and generate probe and conjugate beams by FWM. Conservation of energy requires the simultaneous generation of probe and conjugate.  This leads to quantum correlations between the amplitudes of the beams that make it possible to obtain noise levels below the SNL, or squeezing, when performing intensity difference measurements. In particular, with our source we are able to generate twin beams with 9~dB of intensity difference squeezing (equivalent to noise levels~87\% below the SNL).

\begin{figure}[hbt]
\centering
\includegraphics{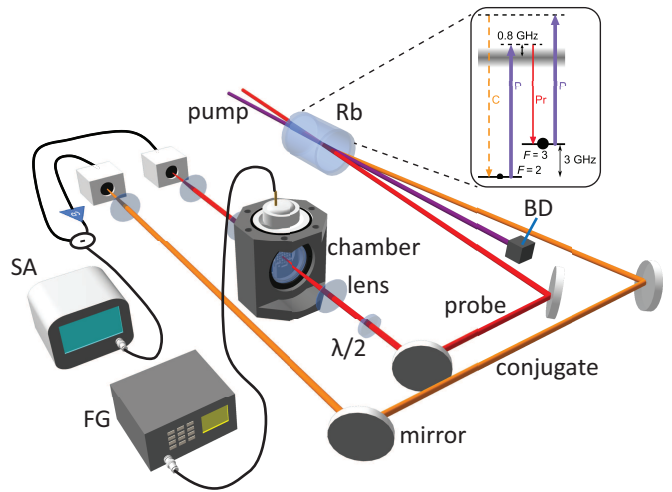}
\caption{Schematic of the experimental setup. One of the twin beams (probe) generated with a four-wave-mixing process is sent through a plasmonic sensor inside a chamber, while the other one (conjugate) acts as a reference for  intensity difference noise measurements. The set up is used to detect small changes in the refractive index of air with a sensitivity below the shot-noise limit. Inset: double-$\Lambda$ energy level scheme on which the FWM process is based. BD: beam dump, SA: spectrum analyzer, FG: function generator.}
\label{fig:experiment}
\end{figure}

\begin{figure*}[hbt]
\centering
\includegraphics{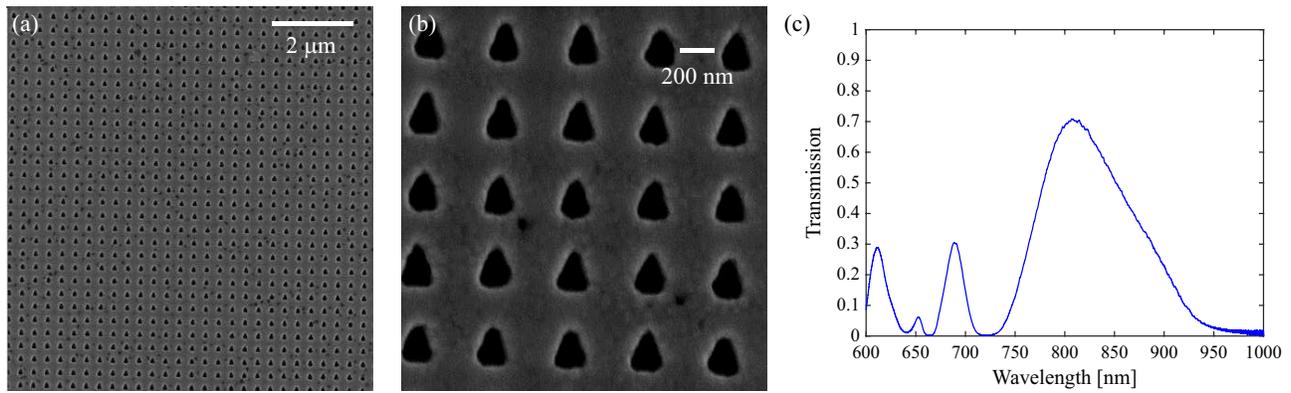}
\caption{Plasmonic structures used as sensor. (\textbf{a}) and (\textbf{b}) are SEM images of the triangular nano-hole array used as the plasmonic sensor. (\textbf{c}) Transmission spectrum of the plasmonic structure measured with a white light source.  At the probing wavelength of 795~nm the EOT transmission is $\sim$66\%.}
\label{fig:sensor}
\end{figure*}
The plasmonic structure that serves as a sensor consists of an array of isosceles triangular  sub-wavelength nanoholes (base 230~nm, side 320~nm, and pitch 400~nm) patterned by electron beam lithography in a 100 nm thick Ag film evaporated on an ITO-coated boro-aluminosilicate glass substrate, as shown in Figs.~2(a,b). The plasmonic structure has an overall size of 200~$\mu$m$\times$200~$\mu$m and the nano-holes are arranged in a square grid. This configuration leads to an EOT transmission of $\sim$66\% at the probing wavelength of 795~nm, as shown in Fig.~2(c). A layer of PMMA is deposited on top to protect the Ag from oxidizing. We have verified that this layer of PMMA does not significantly affect the functionality or sensitivity of the sensor for the ultrasound-based measurements described here. The probe beam is focused to a waist diameter~$<20~\mu$m into the plasmonic structure to avoid beam diffraction from the edges. Given that the plasmonic structures are polarization dependent, a~$\lambda/2$~wave-plate is placed before the sensor in the experimental setup, as shown in Fig.~1, to maximize the transmission.

After the FWM process, the probe beam is used for probing the plasmonic sensor, which is placed inside a hermetically sealed chamber, while the conjugate beam acts as a reference for an intensity-difference measurement. Probe and conjugate beams are detected by two independent photo-detectors and the resulting signals are subtracted with a hybrid junction. The power spectrum of the difference signal is then measured with a spectrum analyzer. The power of the probe beam after the plasmonic sensor is stabilized and kept at 70~$\mu W$ for all the measurements to avoid saturation of the photo-detectors.

In order to study the response of the plasmonic sensor to changes in the index of refraction of the air, the plasmonic structure is placed inside a chamber that provides a well controlled and stable environment. An ultrasound transducer is used to introduce pressure waves inside the chamber that lead to a modulation of the index of refraction of the air around the sensor at the driving frequency of the transducer. The average modulation amplitude of the refractive index ($\Delta n$) along the path of the probe beam inside the chamber as a function of the driving voltage of the transducer is calibrated by placing the chamber in one of the arms of a Mach-Zehnder interferometer (see supplementary information).

To compensate for the losses introduced by the plasmonic structure and other optical elements on the probe, the photodetector for the conjugate has an adjustable electronic gain that allows us to obtain the largest possible noise reduction when performing the differential measurements.
For our current configuration, the squeezing level of 9~dB initially present in the twin beams is reduced to 4~dB (60$\%$ below the SNL) after probing the plasmonic sensor and optimizing the electronic gain. A measure of the SNL is obtained by considering the corresponding classical configuration in which the same detection optimized for the twin beams is used, but with classical coherent states of the same power as the probe and the conjugate.

\section{Results and discussion}

\begin{figure*}[hbt]
\centering
\includegraphics{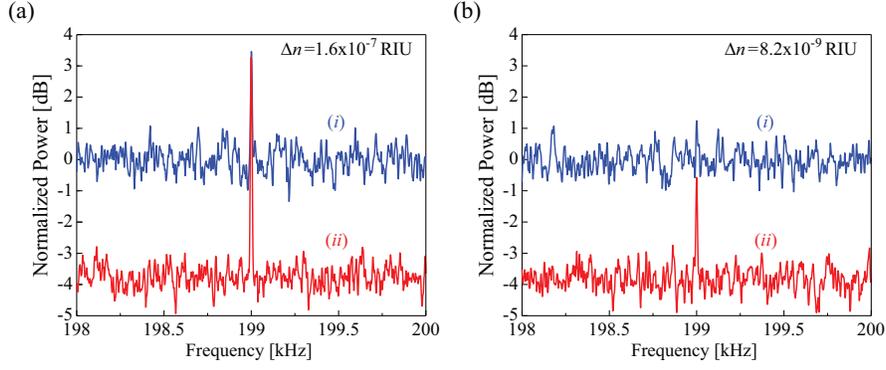}
\caption{Enhancement of SNR with twin beams. (\textbf{a}) Measured power spectra when probing the sensor with coherent states, trace $(i)$, and with twin beams, trace $(ii)$, for a modulation of the air refractive index of $1.6\times 10^{-7}$~RIU. (\textbf{b}) Measured power spectra when the modulation is reduced to $8.2\times 10^{-9}$~RIU. In this case the signal is hidden in the noise when probing with coherent states, trace $(i)$, while it can be resolved when probing with the twin beams, trace $(ii)$. Settings for the spectrum analyzer: Resolution bandwidth (RBW)=10~Hz, Video bandwidth (VBW)=1 Hz, center frequency=199~kHz, span=2 kHz. All traces are averaged 50 times.}
\label{fig:span}
\end{figure*}
When the ultrasonic transducer is driven at its resonant frequency of 199~kHz, the resulting modulation in the index of refraction leads to a modulation in the transmission of the probe beam through the sensor. Figure~3 shows the measured power spectrum of this signal normalized to the SNL. As can be seen in Fig.~3(a), for a modulation of $\Delta n =1.6\times 10^{-7}$~RIU, the signal is resolved with both coherent states and twin beams. However, an enhancement in the signal-to-noise ratio (SNR) is obtained with the twin beams due to the 4~dB reduction in noise. This makes it possible to resolve smaller modulations, as can be seen in Fig.~3(b) where a modulation of $\Delta n =8.2\times 10^{-9}$~RIU is resolvable only with the twin beams.

To obtain a measure of the sensitivity of the plasmonic sensor and the enhancement when using quantum states of light, we perform measurements in which the driving voltage of the ultrasonic transducer is linearly ramped as a function of time. Figure~4(a) shows the measured signal at the driving frequency of the transducer for both coherent states and twin beams. As the driving voltage decreases, the modulation signal decreases until it reaches the noise floor (horizontal traces). At this point it is no longer possible to detect the modulation in the index of refraction as the measured signal is dominated by noise. The noise floor is determined by the noise of the optical field used to probe the sensor and is obtained by turning off the refractive index modulation. As can be seen, the noise starts to dominate the measured signal for lower values of $\Delta n$ when probing the sensor with the twin beams. An accurate measure of $\Delta n_{\rm min}$ for each case is obtained by using the data in Fig.~4(a) to calculate the SNR, shown in Fig.~4(b), and determining the 99\% confidence level at which we can distinguish the signal from the noise (see supplementary information).  Taking the detection bandwidth of 100 Hz into account, which is set by the resolution bandwidth (RBW) of the spectrum analyzer, the sensitivity when probing with coherent states is $\sim8.6\times 10^{-10}$~RIU/$\sqrt{\rm Hz}$, while this value is reduced to $\sim5.5\times 10^{-10}$~RIU/$\sqrt{\rm Hz}$ when probing with the twin beams. This translates to a 56$\%$ quantum enhancement, consistent with the measured 4~dB of squeezing when probing the plasmonic sensor (see supplementary information).
\begin{figure*}[hbt]
\centering
\includegraphics{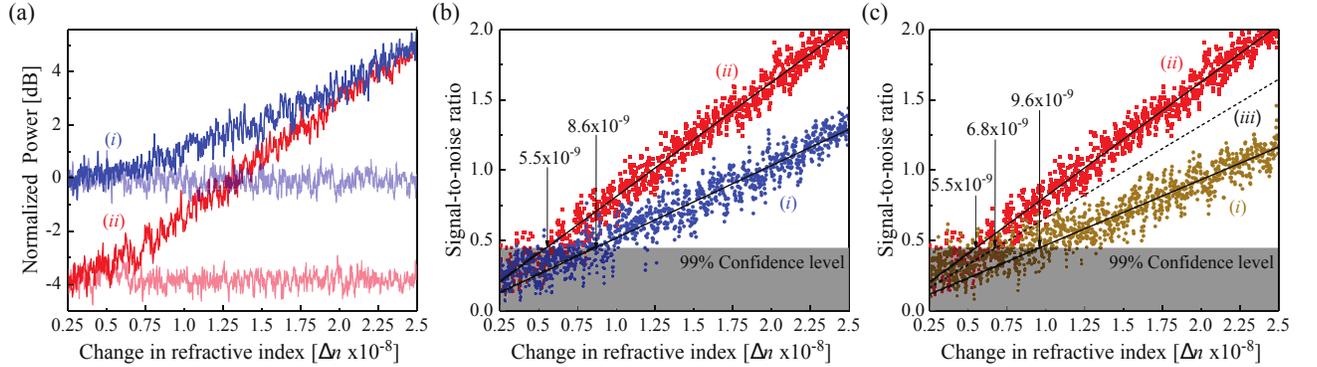}
\caption{Sensitivity enhancement of plasmonic sensors with quantum resources. (\textbf{a}) Measured signal while linearly ramping the driving voltage of the ultrasonic transducer, i.e. increasing the change of refractive index of air ($\Delta n$), when probing with coherent states, trace $(i)$, and with twin beams, trace $(ii)$. The baseline noise for both cases is given by the horizontal traces and was measured by turning off the modulation. All traces are averaged 50 times. Settings for the spectrum analyzer: RBW=100~Hz, VBW=10~Hz, zero span, center frequency=199~kHz, and sweep time=10~s.  (\textbf{b}) Comparison of the SNR when probing the plasmonic sensor with coherent states, trace $(i)$, and with twin beams, trace $(ii)$. A linear fit is used for the calculated SNR. A 99\% confidence bound is used to determine when the signal can be distinguished from the noise and to estimate $\Delta n_{\rm min}$. (\textbf{c}) Comparison with the optimal classical single coherent state configuration.  SNR when probing with twin beams, trace $(ii)$, and with a balanced configuration with two coherent states, each with the same power as the probe, trace $(i)$.  Trace $(iii)$ gives the estimated SNR for the single coherent state configuration.}
\label{fig:Ramp}
\end{figure*}

The quantum-based sensitivity enhancement is mainly limited by the losses after the FWM process since the level of squeezing is degraded by such losses. In our experiment, the major sources of loss are the EOT through the plasmonic sensor ($\sim$34\% loss) and optical losses in the system ($\sim$27\% loss), both of which can be minimized.  The losses in a plasmonic structure can be reduced by optimizing the design and fabrication process. Since the thickness of the metallic film (100 nm) is much smaller than the plasmon propagation length, in principle it is possible to obtain EOT transmission  approaching~100\%~\cite{90Extraordinary,McPeak}. Eliminating the losses in the system would allow us to take full advantage of the~9~dB squeezing, which would result in a quantum-enhancement of~182\% with respect to the corresponding classical configuration (see supplementary information).

It is worth noting that the sensitivities we were able to obtain here are truly state-of-the-art compared with all known classical and quantum plasmonic sensors~\cite{Wang,Kalashnikov,Pooser}. When compared with previous work on quantum-enhanced plasmonic sensors, our results represent an increase in sensitivity of nearly 5 orders of magnitude. In Ref.~\cite{Pooser}, the smallest detectable change of refractive index using quantum resources was 0.001~RIU for a bandwidth of 1~kHz, equivalent to $\sim3\times 10^{-5}$~RIU/$\sqrt{\rm Hz}$, while in Ref.~\cite{Kalashnikov} the smallest detectable change of refractive index was 0.014~RIU with each pixel of their spectra acquired in a 400 second period, which corresponds to a sensitivity of $\sim0.28$~RIU/$\sqrt{\rm Hz}$. Moreover, this sensitivity is comparable to the best previously reported classical sensitivity of the order of $10^{-10}$~RIU/$\sqrt{\rm Hz}$, where optical powers 14 times larger than those used here were used to improve the limits of detection at the expense of potential thermoplasmonic effects.

If we consider the resources used to estimate $\Delta n_{\rm min}$ to be the number of photons used to probe the sensor, then a better classical strategy would be to use a single coherent-state beam.  While in practice it is hard to eliminate all the technical noise from a beam of light, it is still useful to compare to the single-beam strategy. To do so we perform a balanced measurement with two coherent states, each with a power equal  to the probe beam. In this case, $\Delta n_{\rm min}=9.6\times 10^{-10}$~RIU/$\sqrt{\rm Hz}$, as shown by trace $(i)$ in Fig.~4(c). We then estimate the single coherent state sensitivity by taking into account the fact that the balanced coherent state configuration has a noise floor that is twice as large as a single coherent state configuration.  This leads to trace $(iii)$ in Fig.~4(c), which allows us to estimate $\Delta n_{\rm min}=6.8\times 10^{-10}$~RIU/$\sqrt{\rm Hz}$ with a single coherent state of the same power as the probe beam, consistent with the expected sensitivity for our structure (see supplementary information). Even when compared with the optimal classical configuration, the sensitivity is enhanced by 24$\%$ through the use of the twin beams.

Although, in the current measurement configuration we are detecting refractive index changes induced by ultrasound waves, the resulting enhancements can be extended to other sensing applications of plasmonic sensors where the refractive index changes are of the same order of magnitude as the ones presented here.

\section{Conclusion}

In conclusion, we have shown that the use of quantum states of light can improve the detection limits of SPR sensors beyond the SNL. We have demonstrated a refractive index sensitivity comparable to the state-of-the-art classical plasmonic sensors with 14~times less optical power and five orders of magnitude higher than previous proof-of-principle quantum-enhanced plasmonic sensors. When probing the sensor with entangled twin beams, an enhancement of 56$\%$ is obtained with respect to the corresponding classical configuration. Even when comparing with an optimal single coherent state configuration, which is hard to achieve experimentally due to classical technical noise, an enhancement is obtained. Such a quantum enhancement allows for sensitivities below the classical shot-noise limit. Given that quantum-enhanced plasmonic sensors, in general, make it possible to detect smaller changes in the surrounding index of refraction, the results presented here pave the way for further improvements in sensing limits for high precision biomedical and biochemical detection schemes.
\\
\\
This work was supported by the W.~M. Keck Foundation

\section{Acknowledgments}
The fabrication of the plasmonic structures was  performed at  Oak  Ridge National Laboratory, operated by UT-Battelle for the U.S. Department  of  Energy  under  contract  no.~DE-AC05-00OR22725. The nanofabrication and electron microscopy were performed at the Center for Nanophase Materials Sciences, which is a DOE Office of Science User Facility.

\section{Supplemental Information}
Here, we present details on the calibration of the refractive index modulation in the chamber, data taking and analysis, expected sensitivity enhancement, and an order of magnitude estimate of the sensitivity.

\subsection{Calibration of Chamber}
To provide a controlled environment for the characterization of the plasmonic sensor, we place it inside a hermetically sealed chamber. An ultrasonic transducer inside the chamber is driven at its resonance frequency of 199~kHz to generate propagating ultrasonic pressure waves that lead to a modulation of the refractive index of the enclosed air~\cite{Edlen}. To calibrate the amplitude of this modulation ($\Delta n$) as a function of the amplitude of the driving voltage of the transducer, $V_{d}$, we place the chamber in one of the arms of a Mach-Zehnder interferometer. The modulation of the index of refraction introduces a phase shift, which the interferometer converts to an amplitude modulation at its output.

The two outputs of the interferometer are detected with photodiodes and the difference signal is obtained.  This signal is then divided into a low frequency (DC) and a high frequency (RF) component with a bias-tee, which has a cutoff frequency of 100~kHz. The DC component is used to lock the interferometer with a piezo-driven mirror at the zero-crossing of the difference signal, where the signals from the two outputs cancel each other. This provides the most sensitive operational point of the interferometer to changes in phase and makes it possible, for small phase modulations, to make a linear approximation of the difference signal as a function of $\Delta n$.

The RF component contains the signal that results from the index of refraction modulation.  This signal is sent to a spectrum analyzer to measure the magnitude of the intensity modulation in the difference signal as a function of $V_{d}$. This allows us to relate $\Delta n$ at each value of $V_d$ to the measurements performed on the difference signal through the following relation
\begin{equation}\label{interferometer}
  \Delta n=\frac{\lambda B}{\pi A L},
\end{equation}
where $\lambda$ is the wavelength of the light ($795$~nm), $L$ is the propagation length of the beam inside the chamber ($6.35$~mm), $A$ is the amplitude of the difference signal when scanning over a phase shift of $2\pi$, and $B$ is the amplitude of the difference signal due to the modulation of the index of refraction of the air inside the chamber. In order to obtain a reliable calibration, the laser power in the interferometer is stabilized with an acousto-optical modulator (AOM) placed before the interferometer.

From these measurements, a linear relation between the amplitude of the driving signal of the transducer, $V_{d}$, and the change in refractive index, $\Delta n$, is found for the range of modulations that we consider in the experiment. This provides a calibration of the average $\Delta n$ along the path of the beam inside the chamber.

\subsection{Data Analysis}

To obtain a measure of the minimum detectable change in refractive index with the plasmonic sensor, we measure the magnitude of the signal that results from changes in the index of refraction (height of peaks in Fig.~3 of the main text) with the spectrum analyzer as the amplitude of the driving voltage of the ultrasonic transducer is linearly ramped.  These measurements contain a contribution from the signal we are interested in characterizing as well as from the noise due to the probing field.

Due to the noise contribution from the probing field and the logarithmic averaging done by the spectrum analyzer, the estimation of the actual signal from the index of refraction modulation is not trivial, as discussed in Ref.~\cite{Hill1990correction}, and a correction factor has to be considered to estimate the actual signal-to-noise ratio (SNR). Thus, we follow the procedure in Ref.~\cite{Hill1990correction} to estimate the SNR from our measurements as, SNR~(dB)=Measured signal~(dBm) - Measured noise~(dBm) - Correction factor, where the correction factor depends on the SNR. We then convert the SNR to a linear scale as $\sqrt{10^{\text{SNR(dB)}/10}}$. The results from these calculations are shown in Figs.~4(b) and 4(c) of the main text. After this procedure, we find a linear relation between the SNR and the change in refractive index, $\Delta n$, as shown in Figs.~4(b) and 4(c). The value of $\Delta n_{\rm min}$ is then obtained by finding the value at which the linear fit of the SNR can be distinguished with 99\% confidence from the noise.

\subsection{Quantum Enhancement}

As discussed in the main text, the sensitivity of a plasmonic sensor is defined as the minimum resolvable change in the refractive index, which can be expressed as
\begin{equation}\label{sensitivity}
  \Delta n_{\rm min}=\frac{1}{|\partial T/\partial n|}\frac{1}{\sqrt{N}}\sqrt{\frac{\langle(\Delta I)^2\rangle}{|\partial I/\partial T|^2}},
\end{equation}
where we have introduced the effect of averaging $N$ times on the sensitivity.

Once the design of the sensor and power of the beam probing the sensor have been optimized, the use of nonclassical resources offers the only option for further sensitivity enhancements.
We define the level of enhancement obtained when probing with the twin beams as
\begin{equation}
      \frac{\Delta n_{\rm min}^{\rm SNL}-\Delta n_{\rm min}^{Q}}{\Delta n_{\rm min}^{Q}}=\sqrt{\frac{\langle(\Delta I)^2_{SNL}\rangle}{\langle(\Delta I)^2_{Q}\rangle}}-1=\sqrt{{\frac{1}{R}}}-1,
\end{equation}
where $\Delta n_{\rm min}^{\rm SNL}$ and $\Delta n_{\rm min}^{Q}$ are the sensitivities when probing the sensor with classical resources or quantum states, respectively, and $R\equiv\langle(\Delta I)^2_{Q}\rangle / \langle(\Delta I)^2_{SNL}\rangle$ is a measure of the normalized noise of the input probing field or level of squeezing in linear scale. Here $\langle(\Delta I)^2_{Q}\rangle$ and $\langle(\Delta I)^2_{SNL}\rangle$ are the noise variance for quantum and coherent states, respectively. As a result, the amount of enhancement depends only on the amount of noise reduction, or squeezing, due to the quantum states of light.

Given the 4~dB of squeezing ($R=0.4$) that remains after probing the plasmonic sensor with the twin beams, we expect an enhancement by 58$\%$ over the corresponding classical configuration.  This level is consistent with the measured enhancement. Once we are able to minimize the sources of loss we will be able to take full of advantage of the 9~dB of squeezing ($R=0.126$) present in the twin beams, which would lead to an enhancement of 182$\%$.

\subsection{Order of Magnitude Sensitivity Estimation}

We can obtain an order of magnitude estimate of the sensitivity of the plasmonic sensor used in our experiment by considering the configuration in which a single coherent state beam is used to probe the sensor.

First, we consider the contribution to the sensitivity from the plasmonic structure, that is, the term $|\partial T/\partial n|=|\partial T/\partial \lambda|\times|\partial\lambda/\partial n|$ in Eq.~(\ref{sensitivity}). Here, $|\partial T/\partial \lambda|$ corresponds to the slope of the transmission spectrum as a function of wavelength and can be directly obtained from the white light spectrum shown in Fig.~2(c). From here we obtain $|\partial T / \partial \lambda| \sim 0.006$~nm$^{-1}$ at $\lambda=795$~nm with $T=0.66$. For the $|\partial \lambda/\partial n|$ term, we can obtain an estimate  following the relationship from the Ref.~\cite{eftekhari2008polarization}
\begin{equation}
S=\frac{\Delta \lambda}{\Delta n}=\frac{d}{\sqrt{{p^2+q^2}}}\sqrt{\left(\frac{\varepsilon_m}{n^2+\varepsilon_m}\right)^3},
\end{equation}
where $d$ is the periodicity of the nanohole structure, $n$ is the refractive index of the medium around the senors, $\varepsilon_m$ is the permittivity of the metal film, and $p$ and $q$ correspond to the plasmonic mode that is excited. For our plasmonic structure, $d \sim 400$~nm, $n = 1$, and $\varepsilon_m=-24.5+i 1.83$ at $\lambda=795$~nm~\cite{Rakic1998}. This gives an estimate of $S=|\frac{\Delta \lambda}{\Delta n}|\sim 425$~nm/RIU for $p=1$ and $q=0$.  Combining these results, we obtain $|\partial T/\partial n| \sim 2.5~{\rm RIU}^{-1}$.

Next, we consider the term $\sqrt{\langle(\Delta I)^2\rangle/|\partial I/\partial T|^2}$, which depends on the properties of the field used to probe the sensor. Given that the sensor transduces changes in index of refraction to changes in optical transmission, we need to determine how losses affect the properties of the probing field. In the presence of losses, the field operator ($\hat{a}$) transforms according to
\begin{equation}\label{field}
\hat{a}\rightarrow \sqrt{T}\hat{a}+\sqrt{1-T}\hat{a}_{v},
\end{equation}
where $T$ is the intensity transmission and $\hat{a}_{v}$ is the corresponding vacuum mode that couples in as a result of the losses in the system. The mean intensity and variance are then given by
\begin{eqnarray}
    I&=&T\langle\hat{a}{^{\dagger}\hat{a}}\rangle=T\langle{\hat{I}}\rangle_{0}\label{mean}\\
    \langle\Delta{I^2}\rangle&=&T^2\langle{{\Delta\hat{I}}^2}\rangle_{0}+T(1-T){\langle{\hat{I}\rangle}}_{0},
\end{eqnarray}
where $\langle\hat{I}\rangle_{0}$ and $\langle{{\Delta\hat{I}}^2}\rangle_{0}$ are the input mean intensity and intensity variance of the field probing the sensor.  For the particular case of a coherent state $\langle{{\Delta\hat{I}}^2}\rangle_{0}=\langle\hat{I}\rangle_{0}$ and
\begin{equation}\label{sensitivity1}
  \sqrt{\frac{\langle(\Delta I)^2\rangle}{|\partial I/\partial T|^2}}=\sqrt{\frac{T}{\langle \hat{I}\rangle_{0}}}.
\end{equation}
For all the measurements, the power of the probing beam, $P$, was stabilized to $70~\mu$W, which translates to a photon flux of $F=P/{h\nu}=P\lambda/(hc)= 2.8\times10^{14}$~photons/sec at $\lambda=795~nm$. Thus for a measurement bandwidth of 1~Hz, ${\langle \hat{I} \rangle}_{0} \sim 1.5\times10^{14}$ photons.

Finally, we need to take into account the effective averaging with $N=500$, which takes into account the ratio of RBW to VBW and the number of traces averaged on the spectrum analyzer, and the detection bandwidth of 100~Hz.
With these numbers, we obtain an order of magnitude estimation of the sensitivity $\Delta n \sim 1\times10^{-9}~{\rm RIU}/\sqrt{\rm Hz}$, consistent with the measured sensitivities for our structure.

While the measurements presented on the manuscript have been performed with a PMMA layer over the plasmonic structure, we have verified that the sensitivities remain of the same order of magnitude after the PMMA layer has been removed.

\end{document}